\documentclass[letter,english]{article}
\usepackage{emulateapj,epsfig,onecolfloat}

\newcommand{\simgt}{\gtrsim}

\begin{document}
\twocolumn
[
\title{Hydrodynamic Simulations of the Sunyaev-Zel'dovich effect(s)}
\author{Volker Springel, Martin White, and Lars Hernquist}
\affil{Harvard-Smithsonian Center for Astrophysics, Cambridge, MA 02138}
\authoremail{vspringe@cfa.harvard.edu}
\authoremail{mwhite@cfa.harvard.edu}
\authoremail{lhernqui@cfa.harvard.edu}

\begin{abstract}
\noindent
\rightskip=0pt We have performed a sequence of high resolution
hydrodynamic simulations of structure formation in a $\Lambda$CDM
model to study the thermal and kinetic Sunyaev-Zel'dovich (SZ)
effects. Including only adiabatic gas physics, we demonstrate that our
simulations for the thermal effect are converged down to sub-arcminute
scales. In this model, the angular power spectrum of CMB anisotropies
induced by the thermal effect peaks at $\ell\simeq 10^4$, and reaches an
amplitude just below current observational upper limits.  Fluctuations
due to the kinetic effect are a factor of $\simeq 30$ lower in power
and peak at slightly smaller angular scales. We identify individual SZ
sources and compute their counts as a function of source strength and
angular size. We present a preliminary investigation of the
consequences of an early epoch of energy injection which tends to
suppress power on small angular scales, while giving rise to
additional power on large scales from the reheated IGM at high
redshift.
\end{abstract}
 ]

\section{Introduction} \label{sec:intro}

On angular scales below $10'$, the microwave sky carries the imprint of
large-scale structure in the low-$z$ universe.  In particular, cosmic
microwave background (CMB) photons propagating through the universe
are inverse Compton or Doppler scattered by hot electrons along their
path, either in dense structures such as clusters of galaxies or more
generally in hot gas in the intergalactic medium.  Inverse Compton
scattering conserves the number of photons but preferentially
increases their energy, leading to a spectral distortion whose
amplitude is proportional to the product of the electron temperature
and density (or pressure).  Doppler scattering induces an intensity
fluctuation with the same spectral shape as the CMB itself.  These
effects were first described by Sunyaev \& Zel'dovich
(\cite{SZ72,SZ80}) and are known as the thermal and kinetic SZ
effects, respectively (for recent reviews see Rephaeli~\cite{Rep} and
Birkinshaw~\cite{Bir}).

The thermal effect is one of the primary sources of secondary
anisotropies in the CMB on small angular scales.  The change in the
(thermodynamic) temperature of the CMB resulting from scattering off
non-relativistic electrons is
\begin{eqnarray}
{\Delta T\over T} &=& \phantom{-2}y \left( x{{\rm e}^x+1\over {\rm e}^x-1}-4 \right) \\
  &\simeq& -2y\qquad \mbox{for }\ x\ll 1\, ,
\end{eqnarray}
where $x=h\nu/kT_{\rm CMB}\simeq \nu/56.85\;$GHz is the dimensionless
frequency, and the second expression is valid in the Rayleigh-Jeans
limit which we shall assume henceforth. The quantity $y$ is known as
the Comptonization parameter and is given by
\begin{equation}
y\equiv \sigma_T\int d\ell\ {n_e k(T_e-T_{\rm CMB})\over m_e c^2}\, ,
\label{eqn:ydef}
\end{equation}
where the integral is performed along the photon path.  Since $T_e\gg
T_{\rm CMB}$ the integrand is proportional to the integrated electron
pressure along the line of sight.

The kinetic SZ effect arises from the motion of ionized gas with
respect to the rest-frame of the CMB. The resulting temperature fluctuation
is ${\Delta T / T} = - b$, where
\begin{equation}
  b\equiv \sigma_T \int d\ell \ n_e {v_r\over c}
\end{equation}
measures the magnitude of the effect along the line of sight if $v_r$
is the radial peculiar velocity of the gas.  The different dependence
on frequency of the two effects can, in principle, be used to
disentangle them observationally, though the primary CMB anisotropies
provide ``noise'' for the kinetic SZ signal.  The imprint of the SZ
effects on the CMB has been studied by a number of authors (Ostriker
\& Vishniac \cite{OsVi}, Persi et~al.~\cite{PSCO}, da Silva
et~al.~\cite{SBLT}, Refregier et~al.~\cite{RKSP}, Seljak
et~al.~\cite{SelBurPen}, among others), but the theoretical
predictions still exhibit substantial quantitative uncertainty.

In this paper, we report first results of a sequence of high
resolution hydrodynamic simulations designed to study the SZ signal on
small angular scales.  Our simulations can be used to calibrate and
test the semi-analytic modelling that has been done before, explore
the effect of varying the physics included, and provide predictions
for upcoming experiments. In Section 2, we describe our simulation set
and we discuss the techniques we use to compute SZ maps.  In Section
3, we present our results for the power spectra of thermal and kinetic
SZ effects, and we show SZ source counts as a function of source
strength and source size. Finally, we summarize and discuss our
results in Section 4.

\section{Method} \label{sec:method}

\begin{figure}[t]
\resizebox{3.7in}{!}{\includegraphics{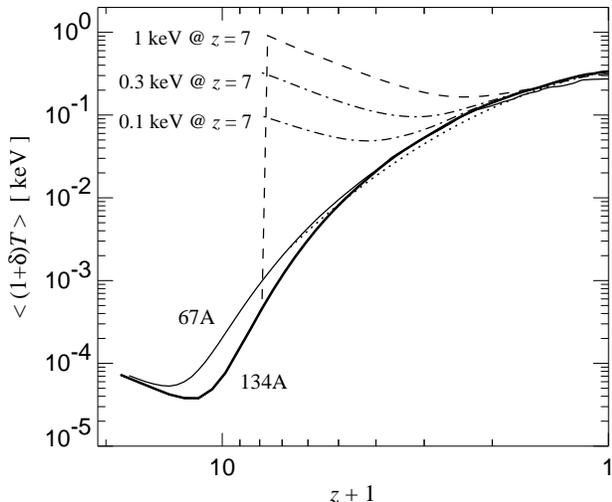}}
\vspace*{-1.2cm}\ \\%
\caption{\footnotesize  The evolution of the density weighted temperature with redshift.
The solid and dashed lines are the simulation results, while the dotted
line shows our predictions based on the Press-Schechter theory.\vspace*{-0.3cm}}
\label{fig:tvsz}
\end{figure}

We are interested primarily in the consequences of assumptions about
the physical state of the gas for predictions of the SZ signal, so we
have focussed on one particular cosmological model, chosen from the
$\Lambda$CDM family.  For definiteness, we have adopted the
``concordance'' model of Ostriker \& Steinhardt~(\cite{OstSte}) which
has $\Omega_{\rm m}=0.3$, $\Omega_\Lambda=0.7$, $H_0=100\,h\,{\rm
km}\,{\rm s}^{-1}{\rm Mpc}^{-1}$ with $h=0.67$, $\Omega_{\rm B}=0.04$,
$n=1$ and $\sigma_8=0.9$ (corresponding to $\delta_H=5.02\times
10^{-5}$).  This model yields a reasonable fit to the current suite of
cosmological constraints and as such provides a good framework for
making realistic predictions.

We ran a number of simulations on the Cray T3E at the San Diego
Supercomputing Center using the {\small GADGET\/} Tree/SPH code
(Springel, Yoshida \& White~\cite{SprYosWhi}).  Each simulation
employed $2\times 224^3$ particles and was carried out on 64
processors.  Our basic model had a box size of $134\,h^{-1}$Mpc, so
the dark matter particles had masses of $1.5\times
10^{10}\,h^{-1}M_\odot$ and the SPH particles $2.4\times
10^{9}\,h^{-1}M_\odot$.  The SPH densities were computed from 32
neighbours, so our minimum gas resolution was roughly $8\times
10^{10}\,h^{-1}M_\odot$. Our base simulation (134A) was evolved from
$z=50$ to $z=0$, and the gravitational force softening was of a spline
form (e.g.~Hernquist \& Katz~\cite{HeKa}), with a
Plummer-equivalent softening length of $25\,h^{-1}{\rm kpc}$
comoving.

As a test of our numerical resolution we also simulated the same
cosmological model in a $67 h^{-1}\,$Mpc box (evolved from $z=100$ to
the present; 67A), with a softening length half that of the 134A run
and eight times better mass resolution.  In all our runs, the full
simulation box was output at 333 different times, between redshifts
$z\simeq 19$ and $z=0$.  The simulations reported here followed only
adiabatic gas physics.  Since the SZ effect is dominated by massive
objects with large cooling times this is a reasonable first
approximation.

To investigate the effect of an early epoch of energy injection we
took the $z=7$ output of 134A and added $0.1\,$keV, $0.3\,$keV or
$1\,$keV of energy per (gas) particle.  This injection scenario was
deliberately chosen to be somewhat artificial. A more realistic
approach might be to add the energy only to gas within halos above a
certain density, to describe energy input from galaxy winds or QSO
activity for example, or to add it over an extended period of time.
Unfortunately, the very large parameter space of these possibilities
cannot be explored easily.  We thus decided to examine an ``extreme''
scenario to get a handle on the dependence of the small angle SZ
signal on the amount of energy injected.  The amount of extra energy
was chosen to be plausible given the observed luminosity--temperature
relation in clusters of galaxies (Cavaliere, Menci \&
Tozzi~\cite{CavMenToz}; Wu, Fabian \& Nulsen~\cite{WuFabNul}) and the
expected X-ray background from the diffuse intergalactic medium
(Pen~\cite{Pen1999}).  We added the energy at $z=7$, which we also
took to be the epoch of reionization, so that the energy injection
takes place ``before'' the gas contributes significantly to the SZ
signal.  A visual inspection of slices through the simulation shows
that an energy injection as large as $1\,$keV literally blows gas out
of small halos, leaving only the gas in the larger collapsed objects.
The gas is thus more smoothly distributed in addition to being hotter.
At $z=0$ the slope of the mass weighted temperature--mass relation in
the $1\,$keV simulation is thus changed from a canonical $1.5$ to a
considerably steeper $1.7$.

The evolution of the density weighted temperature in our simulations
is shown in Fig.~\ref{fig:tvsz}.  Note the good convergence between
the large (134A) and the small box (67A) at redshifts $z<5$. In the
67A run, the adiabatic decay of the mean temperature is halted at a
somewhat higher redshift, as expected: due to the
better mass resolution of this simulation, nonlinear structures of
small mass collapsing at earlier times can be resolved.  In the runs
with energy injection, the temperature increases discontinuously at
$z=7$, and then declines again with the adiabatic expansion, until
shock heating takes over at $z\sim 1-3$, depending on the amount of
injected energy.  At $z=0$, all runs yield a mean mass-weighted
temperature of $\simeq 0.3\,{\rm keV}$.

\begin{figure}[t]
\begin{center}
\vspace*{+0.4cm}%
\resizebox{3.3in}{!}{\includegraphics{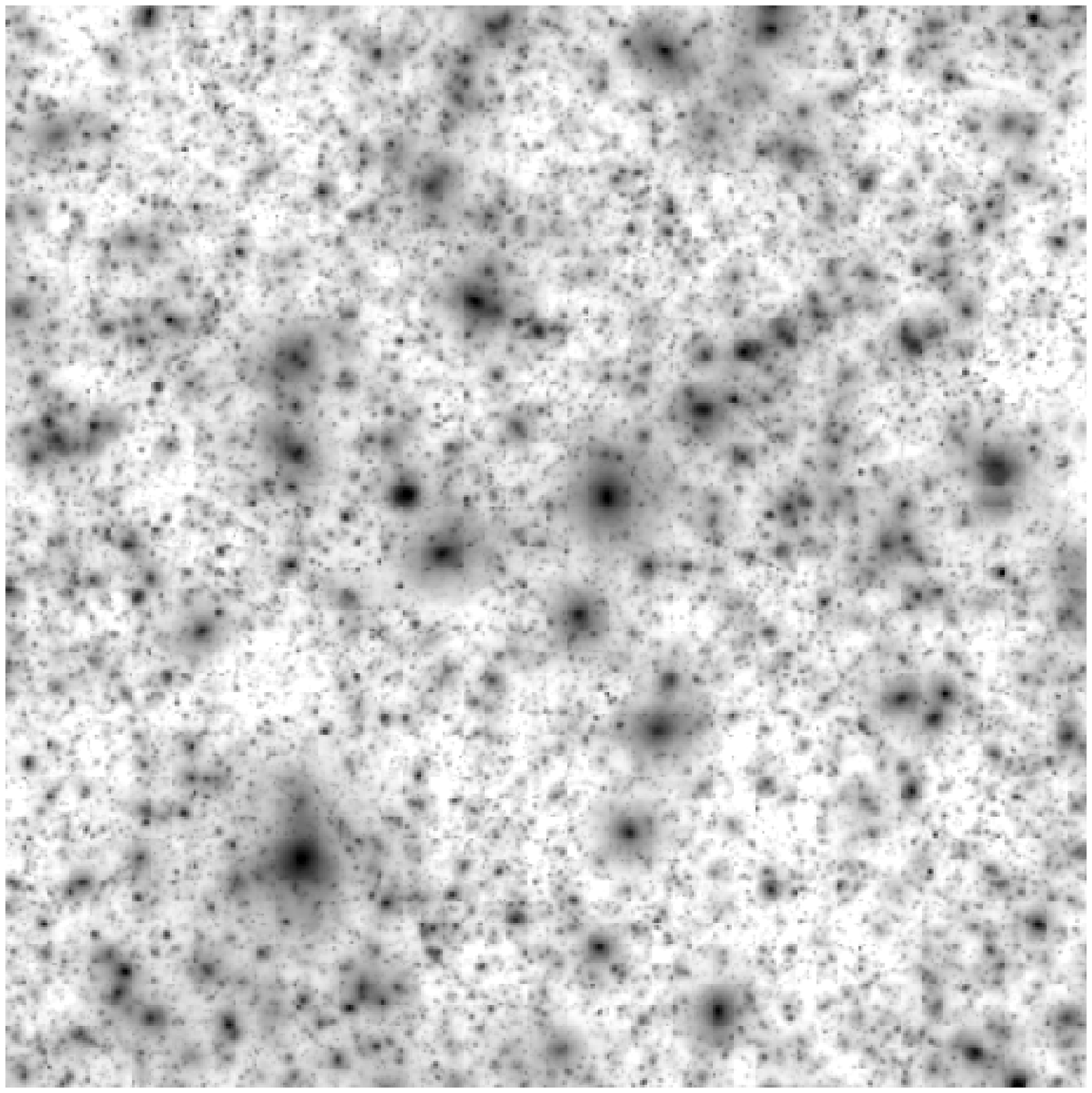}}\vspace*{0.2cm}\\
\resizebox{3.2in}{!}{\includegraphics{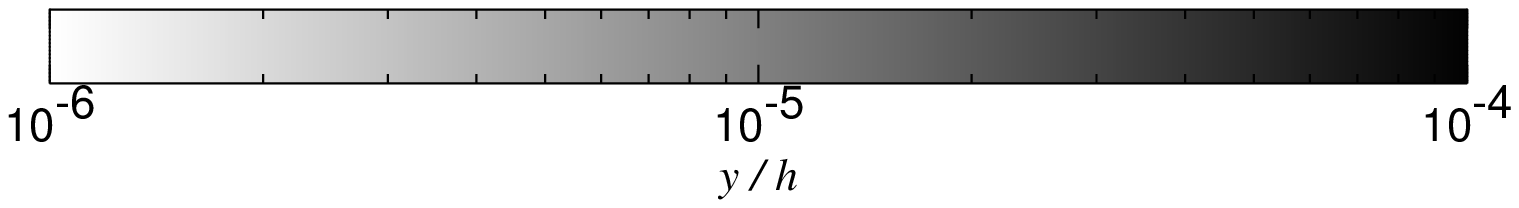}}
\vspace*{0.3cm}
\resizebox{3.3in}{!}{\includegraphics{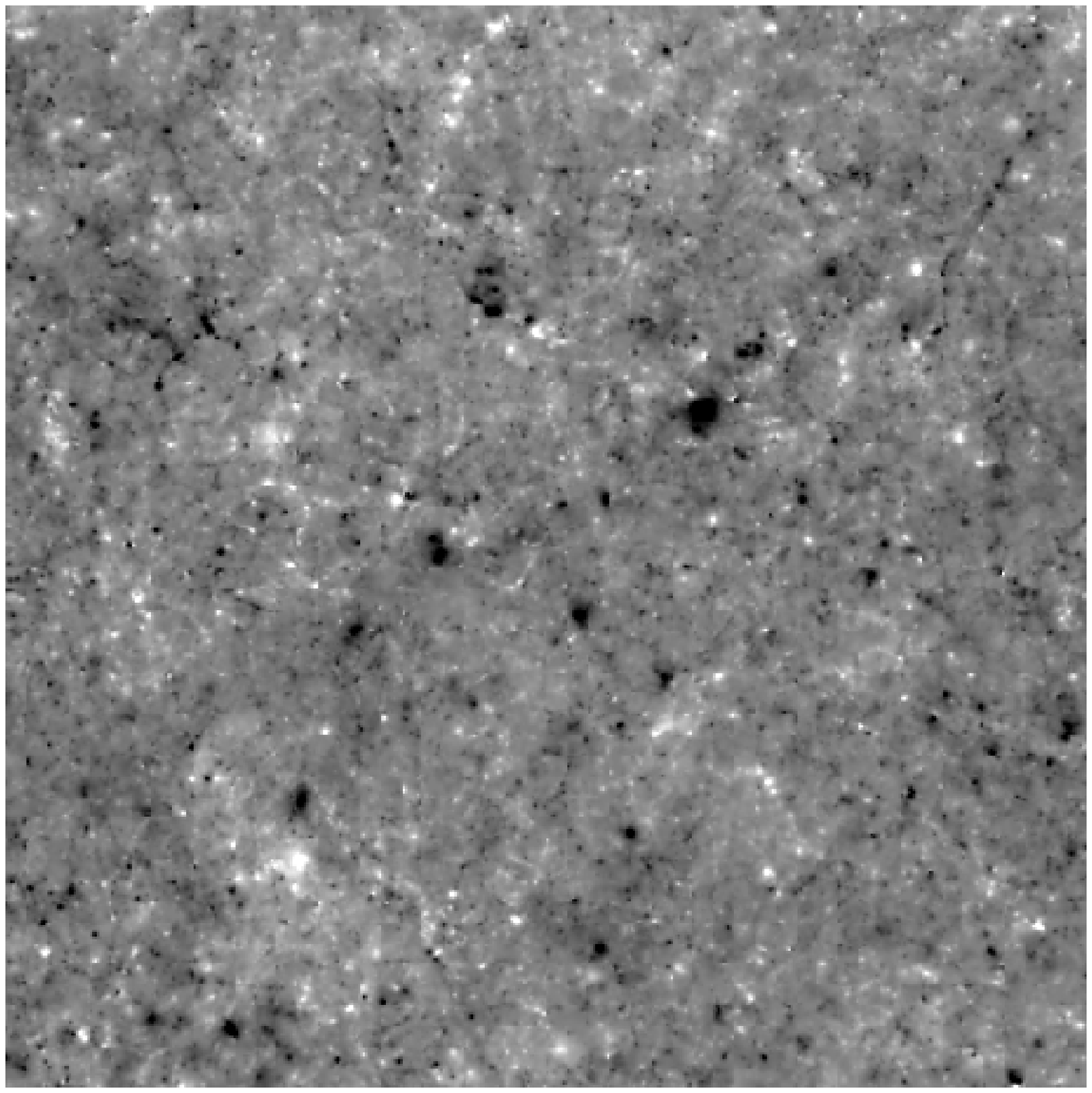}}\vspace*{-0.2cm}\\
\resizebox{1.5in}{!}{\includegraphics{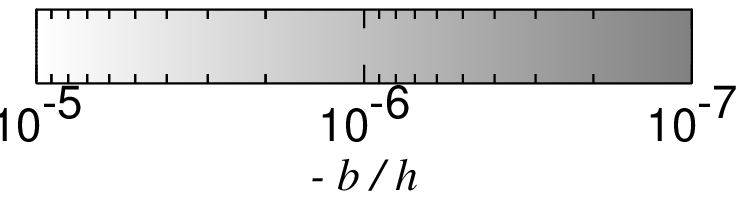}}\ \ \resizebox{1.5in}{!}{\includegraphics{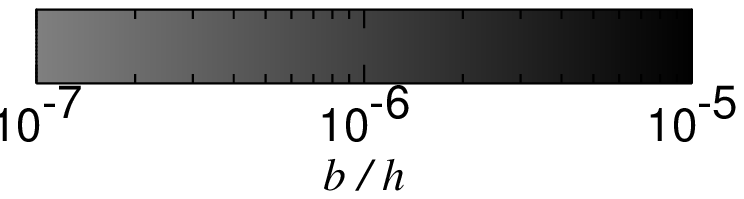}}
\end{center}
\vspace*{-0.4cm}%
\caption{\footnotesize 
Maps of the thermal (top) and kinetic (bottom) SZ effects.  The maps
are $1^\circ$ on a side and cover both the same field of view, here
computed for the 134A simulation.  Notice that unlike the
Comptonization $y$, the kinetic signal $b$ may be negative; we map the
negative values onto the lower half of a logarithmic
grayscale. Structures appearing in white are moving towards the
observer, those in black away.\vspace*{-0.2cm}}
\label{fig:ymap}
\end{figure}

We produced maps in a manner similar to da~Silva et al.~(\cite{SBLT}).
First, the various outputs from the boxes were ``stacked'' back along
the line of sight, with each box randomly translated and oriented in
the direction of either the $x$, $y$ or $z$ axis. Then, a grid of
$512^2$ rays subtending a constant angle of $1^{\rm o}$ from the
observer was traced through the boxes starting at $z=19$.  For our
choice of an opening angle of $1^{\rm o}$, the full $134\,h^{-1}{\rm
Mpc}$ box just subtends an angle equal to the field of view at
$z=19$. For the smaller $67\,h^{-1}{\rm Mpc}$ run, the field of view
reaches the box size at a distance $z\simeq 2$, beyond which we
covered it by periodic replication of the box.

We produced maps of the $y$-parameter, the Doppler $b$-parameter, and
the projected gas and dark matter densities.  For example, the $y$
temperature decrement along any ray was calculated by distributing the
product of pressure and specific volume of the gas particles over the
angular grid:
\begin{equation}
 y_{ij} L_{\rm pix}^2 =
 {\sigma_T\over m_e c^2}\sum_\alpha p_{\alpha} w_{\alpha,ij} \,\, ,
\end{equation}
where $w_{\alpha,ij}$ is the value of the (projected) smoothing kernel
(normalized to unity for the pixels covered) of particle $\alpha$ at
angular grid position $(ij)$, and $L_{\rm pix}^2$ is the physical area
of a pixel at the particle's distance.

The product of pressure and volume, $p_\alpha$, for each SPH particle
is
\begin{equation}
 p = (\gamma-1)(1-Y_p) m u \mu x_e \, ,
\end{equation}
where $\gamma$ is the ratio of specific heats, $Y_p=0.24$ the
primordial ${}^4$He mass fraction, $m$ the particle mass, $u$ the
internal energy per unit mass, $\mu$ the mean molecular weight and
$x_e$ the ratio of electron and hydrogen number densities.  In the
runs reported here, we do not track $x_e$ dynamically. Instead, we set
it to 1.158 for temperatures above $10^4\,$K, corresponding to full
ionization, and take the gas at lower temperature to be neutral. In
reality, the inclusion of an ionizing background would render this
assumption invalid at low redshift, but the small densities of the
affected gas do not lead to any significant contribution to the SZ
signals. These are thus unaffected by our assumption about the
ionization state of the gas, provided ionization happens late enough.
On the other hand, it is crucial to assume that the gas is neutral at
very high redshift, otherwise the high physical density of the ambient
gas in this regime would give rise to a significant contribution to
the SZ effects. In particular, the finite box size of our simulations
would then lead to a strong signal for the kinetic effect on large
angular scales, reflecting motion of gas due to the largest modes in
the box.

\begin{figure}[t]
\begin{center}
\resizebox{3.5in}{!}{\includegraphics{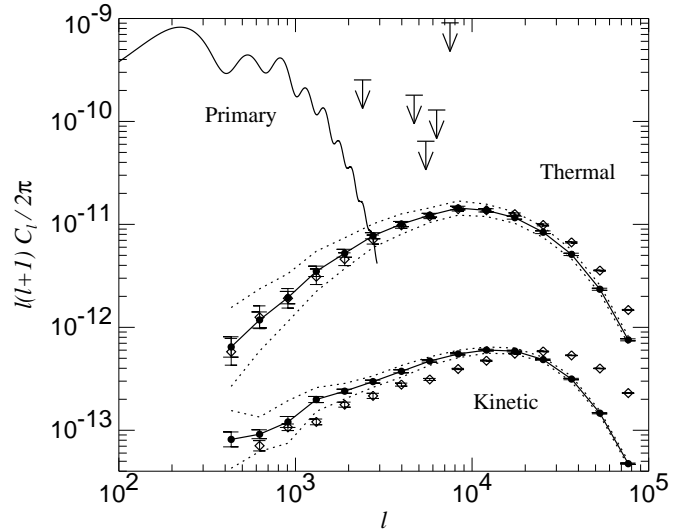}}
\end{center}
\vspace*{-0.7cm}%
\caption{\footnotesize The angular power spectrum of the SZ effect
from an ensemble of 15 maps.  The upper solid line shows the primary
CMB anisotropies for this model. The points indicate the mean and
error on the mean from our ensemble of maps. Filled circles mark
results for the 134A simulation, open diamonds are for the smaller
67A-box. The dotted lines indicate the field-to-field variance between
individual $1^\circ$ maps.  The arrows at upper right are 95\% CL
upper limits from a variety of experiments: SUZIE (Church
et~al.~\cite{Chu97}), ATCA (Subrahmanyan et~al.~\cite{Sub93,Sub98}),
BIMA (Holzapfel et~al.~\cite{Holz2000}), Ryle (Jones~\cite{Jon97}),
and VLA (Partridge et~al.~\cite{Part97}). }
\label{fig:lcl}
\end{figure}

\section{Results}

We show typical examples for our thermal and kinetic SZ maps in
Fig.~\ref{fig:ymap}. These maps were produced by coadding a large
number of partial maps, each giving the contribution of one of the
boxes that we stacked along the photon's path. It is interesting to
note that filamentary structure is easily detected in these partial
maps with their small depth in redshift space. However, in the full
projection along the backward photon path, filaments are largely
hidden in the high level of background arising from the summation over
many of these structures. Things may not be quite so bad for the
kinetic effect, where it seems that a somewhat larger degree of
filamentary structure manages to survive at appreciable signal.  Note
that sources that are ``bright'' in the thermal effect are not
necessarily among the brightest in the kinetic effect, and vice versa.
Often, the kinetic effect shows neighboring large peaks of opposite
sign. This may be useful in strategies for identifying clusters and
superclusters (Diaferio, Sunyaev \& Nusser~\cite{Dia2000}).

\begin{figure}[t]
\begin{center}
\resizebox{3.5in}{!}{\includegraphics{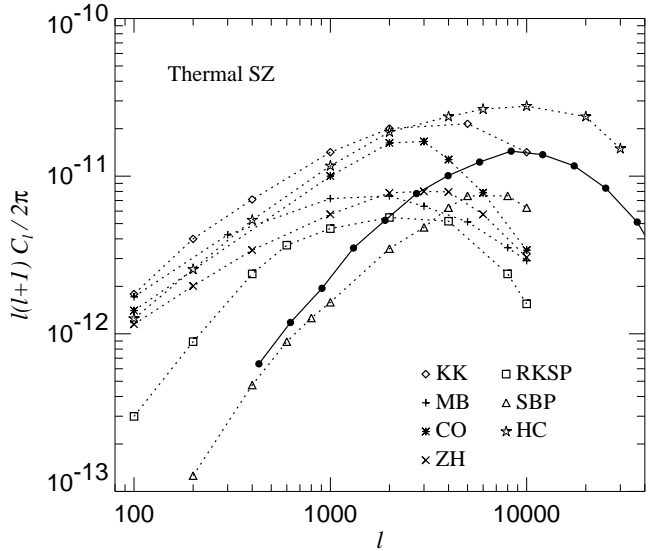}}
\end{center}
\vspace*{-0.4cm}%
\caption{\footnotesize Published estimates of the thermal SZ power
spectrum compared to the result of this work. Filled circles show our
measurement, while other symbols give computations by
the following authors: KK (Komatsu \& Kitayama \cite{Komatsu1999}), MB
(Molnar \& Birkinshaw \cite{Mol2000}), CO (Cooray \cite{Cooray2000}),
ZH (Zhang \& Pen \cite{Zhang2000}), RKSP (Refregier
et~al.~\cite{RKSP}), SBP (Seljak et~al.~\cite{SelBurPen}), and HC
(Holder \& Carlstrom \cite{Hold1999}).}
\label{fig:comp}
\end{figure}

\vspace*{0.3cm}\ \\

\subsection{Power spectra}

For each map we computed a number of statistics, and we in general
averaged the results for 15 random lines of sight to reduce
field-to-field variance.  Of most interest here are the angular power
spectrum of the thermal and kinetic SZ effects (Fig.~\ref{fig:lcl}).

A comparison of our basic $67\,h^{-1}\,{\rm Mpc}$ and
$134\,h^{-1}\,{\rm Mpc}$ runs shows that for the physics we have
included we have converged in the mean mass-weighted temperature and
the angular power spectrum for $\ell\le 20000$, albeit some small
residual resolution effects are visible for the kinetic effect.  The
good level of agreement for the two different runs down to very small
angular scales builds confidence in our numerical technique.  While
this is to be expected as the signal is dominated by structures of
mass $>10^{13}M_\odot$, which should be well covered by our
resolution, it also suggests that finite box size and the box stacking
technique (White \& Hu~\cite{WhWa}) are not a source of significant
uncertainty in our results.

We note that the small box gives slightly more power at $\ell \simgt
20000$ for the thermal effect, indicating its better resolution, while
at lower $\ell$ the convergence is excellent. This is also reflected
in the mean Comptonization, which both models give as $\left<y\right>
=2.6 \times 10^{-6}$.  For the kinetic effect, we observe a higher
level of additional small-scale power for the small box on one hand,
and slightly less power for all $\ell$ below $\ell\simeq 20000$ on the
other hand. This suppression likely originates from a deficit of
large-scale power in the small box, which reduces the amplitude of
peculiar velocities.

Interestingly, the power spectrum of the thermal SZ effect is not far
from current observational upper limits, making it a highly interesting
target for upcoming experiments. The power in fluctuations due to the
kinetic SZ effect is substantially smaller, but may not be out of reach
in future experiments, especially when multi-frequency observations
are employed.

\begin{figure}[t]
\begin{center}
\resizebox{3.5in}{!}{\includegraphics{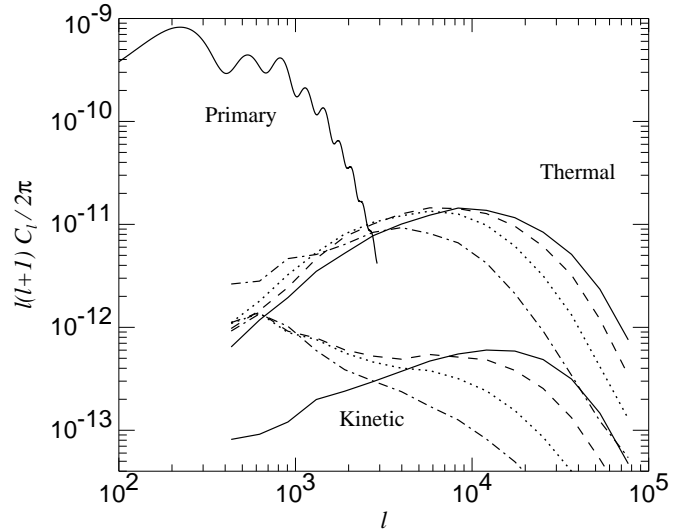}}
\end{center}
\vspace*{-0.7cm}%
\caption{\footnotesize%
Energy injection and its impact on the angular power spectrum of the
SZ effect.  The upper solid line shows the primary CMB anisotropies
for this model, and the lower solid lines give the secondary
anisotropies for thermal and kinetic SZ effects in our 134A
simulation. Dashed, dotted, and dot-dashed lines show the effect of an
injection of 0.1, 0.3, and 1.0 keV of energy at $z=7$, respectively.
Results are based on ensembles of 15 maps for each of our numerical
simulations.}
\label{fig:powerinj}
\end{figure}

It is interesting to compare published results for the power spectrum
of the thermal SZ effect with our measurement, as we have done in
Figure~\ref{fig:comp}. There are order of magnitude differences
between the various results. Note that slight differences in the
adopted cosmological models can partly account for these
discrepancies, but even for cosmological parameters that are very
close, substantial uncertainties remain. One group of results in this
comparison is based on analytical computations, using either
extensions of Press-Schechter theory (Komatsu \& Kitayama
\cite{Komatsu1999}, Holder \& Carlstrom \cite{Hold1999}, Cooray
\cite{Cooray2000b}, Molnar \& Birkinshaw \cite{Mol2000}) or non-linear
perturbation theory (Zhang \& Pen~\cite{Zhang2000}). Compared to our
result, these computations suggest more power on large angular scales,
and their spectrum usually peaks at somewhat larger angular scale,
corresponding to $\ell$ of a few thousand. On the other hand, the
numerical work by Seljak et~al.~(\cite{SelBurPen}), which is based on
simulations using a moving-mesh hydrodynamical code
(Pen~\cite{Pen1998}), suggests a very similar shape for the spectrum,
but with an amplitude which is a factor 1.5-2 lower than our
result. Refregier et~al.~(\cite{RKSP}) use the same simulation
technique as Seljak~et~al.~(\cite{SelBurPen}), albeit at lower
resolution and in combination with a different analysis of the
simulation outputs. Their spectrum peaks at an angular scale of
$\ell\simeq 2000$, roughly at the upper boundary of the author's own
range of confidence of $200<\ell<2000$ for their result.

In general, predictions in the literature for the kinetic SZ effect
show a similar degree of quantitative uncertainty. Note however that
the computations by Bruscoli et~al. (\cite{Brus1999}) are very close
to our result, as is Hu \& White (\cite{HuWh96}) if reionization is
assumed to happen late.  On the other hand, the signals computed for
inhomogeneous reionization (Gruzinov \& Hu \cite{Gruz98}, Benson et~al.
\cite{Bens2000}, Valageas et~al. \cite{Val2000}) are typically an
order of magnitude smaller in power at the peak of the spectrum.

The relatively large scatter in these computations highlights that
there is still substantial quantitative uncertainty in the theoretical
predictions for the SZ power spectrum. High-resolution simulations
like the ones discussed here should help to lead to a more reliable
answer.

An additional complication besides computational uncertainties is that
the theoretical predictions depend quite sensitively on the physics of
reionization and preheating, especially for scales $\ell \simgt 2000$,
where the SZ effect dominates over primary CMB fluctuations. This is
seen when analyzing the SZ signals of our energy injection runs
(Fig.~\ref{fig:powerinj}).  Depending on the amount of energy
injected, fluctuations on small angular scales can be suppressed very
strongly, while additional power on large scales is seen from
relatively diffuse, ionized gas.  This signal on large angular scales
is generated in a small redshift range right after the energy has been
injected.  Note that on the largest scales we probe, the amplitude of
the kinetic SZ effect then becomes dominated by finite box size
effects as the signal reflects the largest modes that are present in
the simulation.

Apart from influencing the power spectrum, energy injection also alters
the mean Comptonization of the models. It becomes $\left<y\right> =
6.9\times 10^{-6}$ for the $0.1\,$keV model, $16.6\times 10^{-6}$ for
the $0.3\,$keV run, and reaches $38.0\times 10^{-6}$ for an injection
of $1\,$keV.  The latter is thus strongly excluded by the upper limit
of $15\times 10^{-6}$ from the COBE FIRAS experiment (Fixsen et
al.~\cite{FIRAS}), and even the $0.3\,$keV model is marginally
excluded.  This highlights the power of the FIRAS constraint.

\begin{figure}[t]
\begin{center}\vspace*{0.2cm}%
\resizebox{3.3in}{!}{\includegraphics{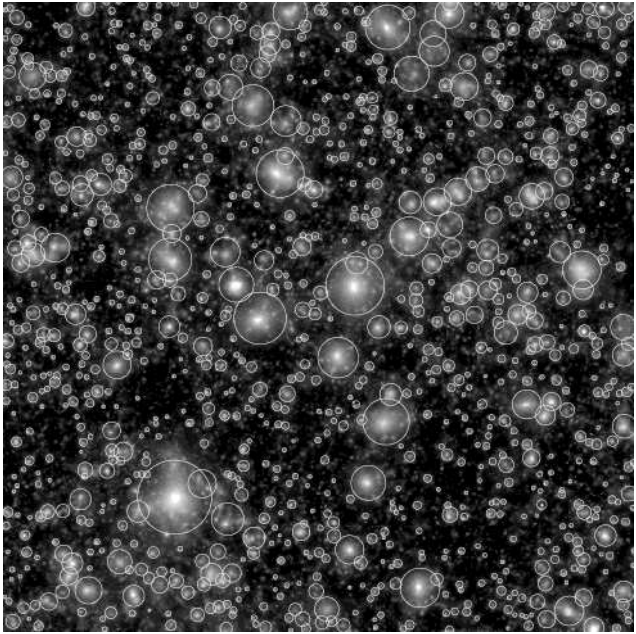}}%
\end{center}%
\vspace*{-0.8cm}\ \\%
\caption{\footnotesize Thermal SZ sources identified with SExtractor.
A circle is drawn for each source, with area equal to the area
determined by the source detection software. The map shown in this
example is the same as in Fig.~\ref{fig:ymap}, but here displayed with
reversed video for graphical clarity.
\vspace*{-0.3cm}}
\label{fig:sourceexample}
\end{figure}

\subsection{Source counts}

The maps for the thermal SZ effect are dominated by discrete sources,
as suggested by visual inspection.  We have used the source detection
software SExtractor (Bertin \& Arnouts \cite{Bertin96}) to count
individual sources, and to measure their strength. A similar technique
has been employed by da~Silva~et~al.~(\cite{SBLT}).
Figure~\ref{fig:sourceexample} shows a typical source detection, here
obtained for the map shown in Fig~\ref{fig:ymap}. To mark each source,
we have drawn a circle with area equal to the area of the ellipse
matched by SExtractor to each source.  We used default settings for
the detection algorithm of the image processing software and let it
estimate background and `noise' levels automatically. Note that our
maps are in principle noise-free, so what is interpreted as noise in
this procedure effectively arises from limitations due to source
confusion, which leads to non-detections of some of the faintest and
smallest sources. In a typical analysis with SExtractor, we resolve
slightly more than half of the total SZ signal into discrete sources.

We measure the strength of a source as the monochromatic brightness change
\begin{equation}
S_\nu = \int_\Omega \Delta B_\nu \, {\rm d}\Omega = f(x) B_{\nu} \int_\Omega y(\theta) \, {\rm d}\Omega
\end{equation}
of the CMB integrated over the solid angle of the source. Here $B_\nu$ is the
Planck spectrum of the primary CMB, and $f(x)$ is the spectral function 
\begin{equation}
f(x)= x \frac{{\rm e}^x}{{\rm e}^x-1} \left( x{{\rm e}^x+1\over {\rm e}^x-1}-4 \right),
\end{equation}
with $x=h\nu/kT_{\rm CMB}$.  In the following, we quote results for a
frequency of 150 GHz. 

\begin{figure}[t]
\resizebox{3.4in}{!}{\includegraphics{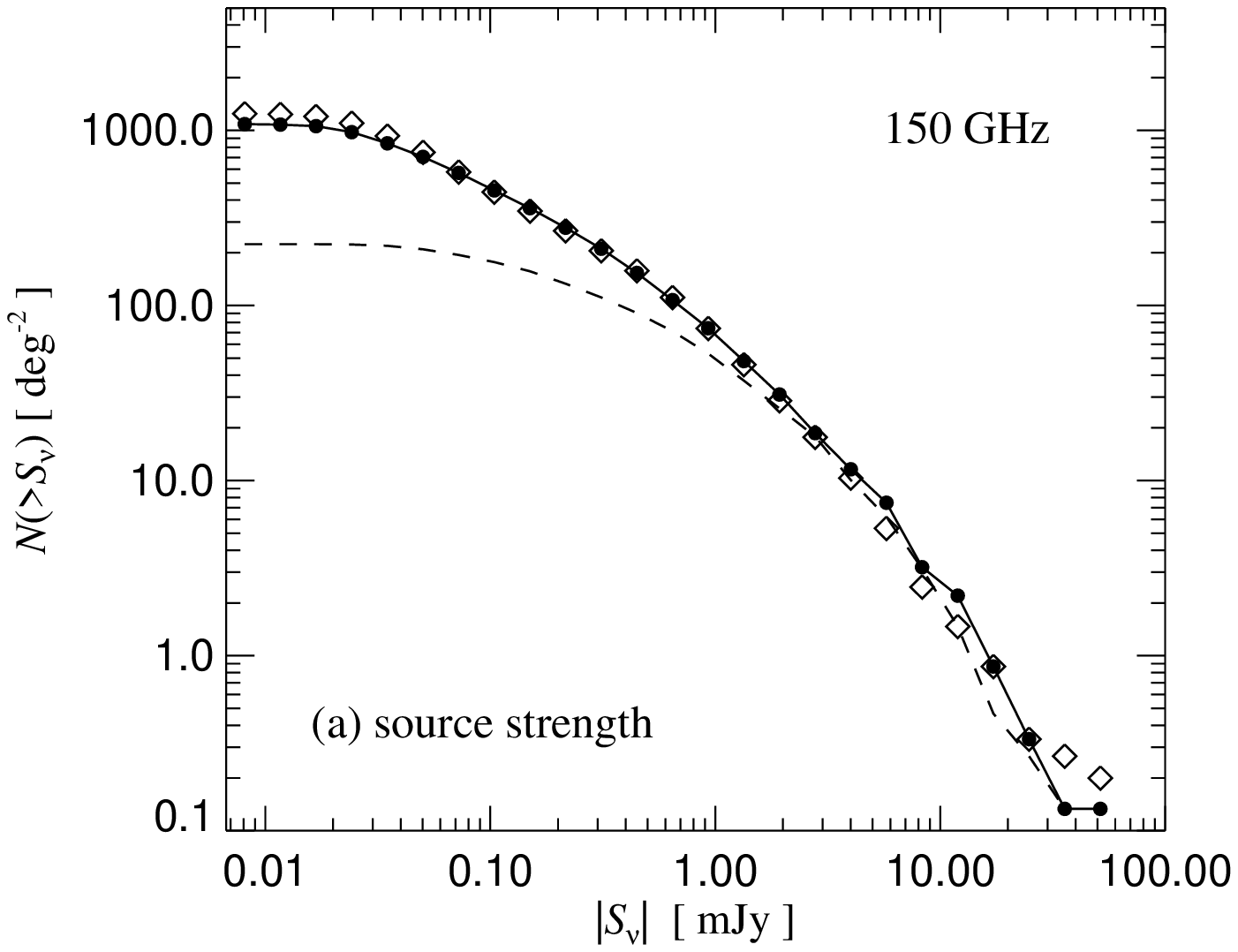}}\\
\resizebox{3.4in}{!}{\includegraphics{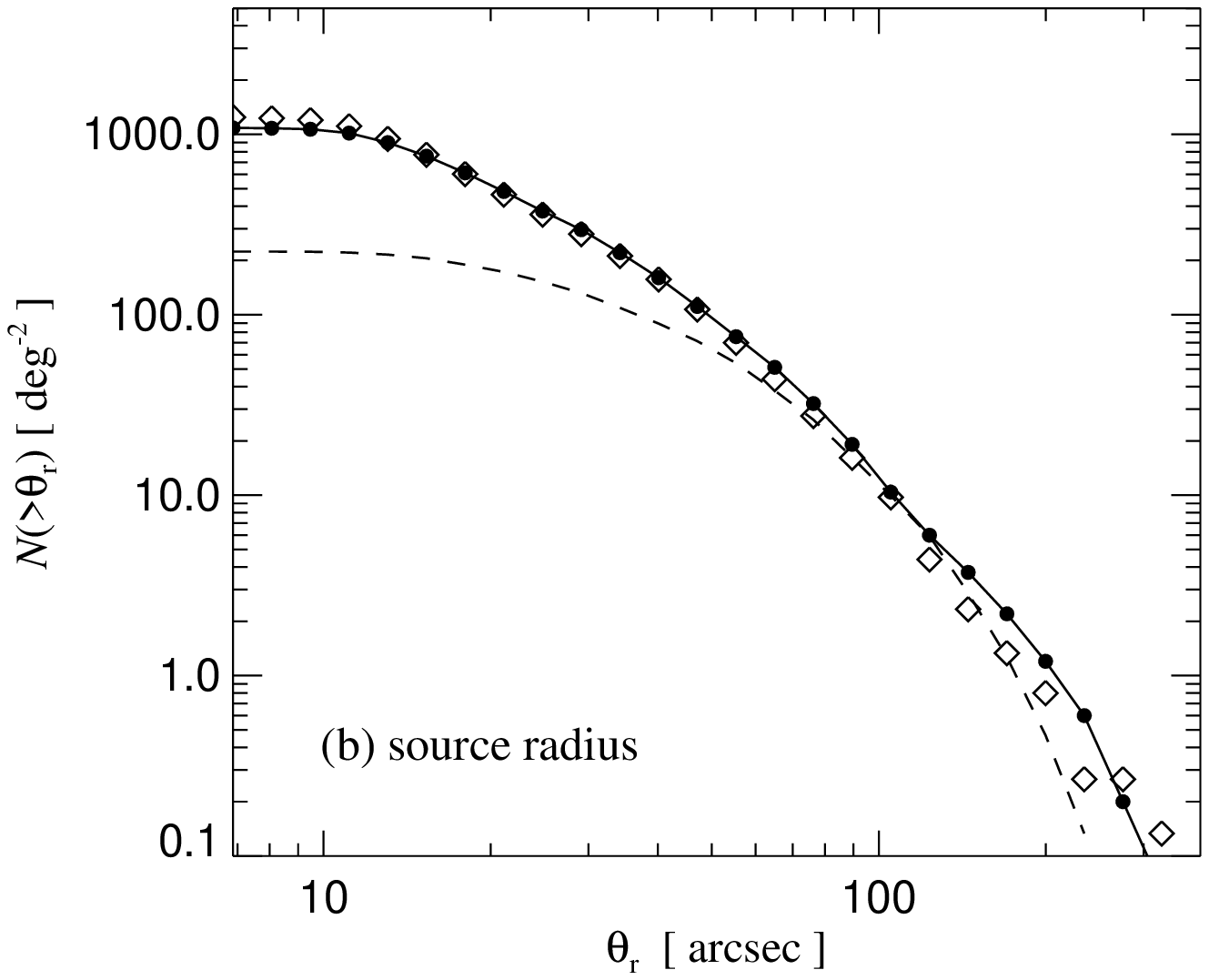}}
\vspace*{-0.2cm}\ \\%
\caption{\footnotesize Cumulative counts of thermal SZ sources per
square degree for models 134A (filled circles) and 67A (diamonds): (a)
as a function of source strength, and (b) as a function of source
radius.  For each source, the source strength $S_\nu$ was defined as
the integrated monochromatic flux decrement, integrated over the solid
angle of the source. The radius of a source was computed based on the
area detected by the source extraction software. In both panels, the
dashed line shows the counts for the run with an injection of $1\,{\rm
keV}$ of energy at $z=7$.
\vspace*{-0.3cm}}
\label{fig:sourcecounts}
\end{figure}

In Fig.~\ref{fig:sourcecounts}, we show the cumulative source counts
per square degree found in the simulations 134A and 67A. We find
remarkably good agreement between the two simulations, which differ by
a factor of 8 in mass resolution. However, the higher resolution run
67A resolves a slightly larger number of faint (small) sources, while
the bigger box 134A shows a small increase of the counts at the bright
(large) end of the distribution. Both of these weak trends reflect
expected effects due to box size and numerical resolution. Notice that
the source counts we find are quite close to predictions by De~Luca
et~al.~(\cite{DeLuca95}) based on Press-Schechter theory, however our
simulations don't show their exact power-law behavior at source
strengths above 1~mJy. The dashed line in Fig.~\ref{fig:sourcecounts}
gives the counts for the simulation with 1~keV energy
injection. Interestingly, the counts for the bright sources are
unaffected, but sources below a flux of 1~mJy (size of $\sim\,$30$''$)
are strongly reduced in number density, and still fainter (smaller)
ones are practically eliminated.

\begin{figure}[t]
\resizebox{3.4in}{!}{\includegraphics{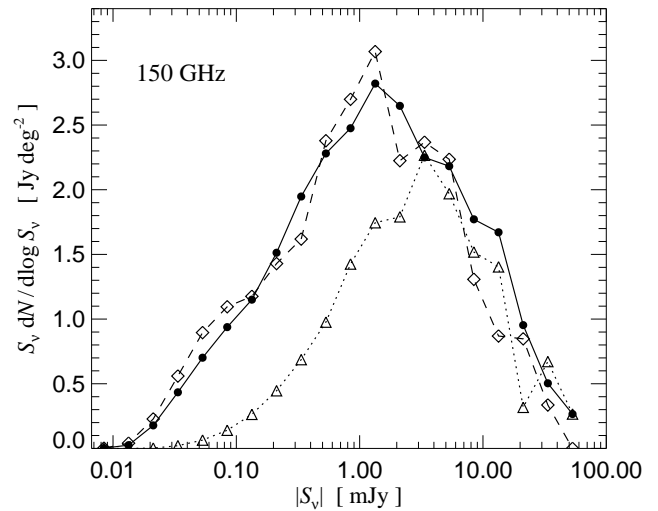}}
\vspace*{-0.2cm}\ \\%
\caption{\footnotesize Differential flux distribution as a function of source
strength for models 134A (filled circles) and 67A (diamonds). For
each source, a source strength $S_\nu$ was defined as the integrated
monochromatic flux decrement, integrated over the solid angle of the
source. Triangles shows flux distribution for the run with an injection of
$1\,{\rm keV}$ of energy at $z=7$.
\vspace*{-0.3cm}}
\label{fig:fluxdist}
\end{figure}

Finally, we plot in Fig.~\ref{fig:fluxdist} the differential flux
distribution as a function of source strength. It is seen that the
bulk of the total flux is provided by sources around 1~mJy, while the
contribution by very faint and very bright sources is negligible.  The
good agreement at the faint end between the two adiabatic runs
suggests that both of these simulations already resolve all the
relevant sources that contribute significantly to the total
Comptonization. Consistent with the results for the source counts, the
energy injection run shows a strong suppression of faint SZ sources,
while the brightest sources are nearly unaffected.

\section{Discussion} \label{sec:conclusions}

We have performed a sequence of high resolution hydrodynamic
simulations of structure formation in a $\Lambda$CDM model to
investigate the thermal and kinetic Sunyaev-Zel'dovich effects.
Figs.~\ref{fig:tvsz}, \ref{fig:lcl}, \ref{fig:sourcecounts}, and
\ref{fig:fluxdist} show that including only adiabatic gas physics our
simulations of the thermal effect are converged down to sub-arcminute
scales. In particular, our results are well converged at the peak of
the power spectrum, and they predict a mean Comptonization of
$\left<y\right> =2.6 \times 10^{-6}$ for the adopted cosmology.

The detection of the brightest SZ sources with a density of $\sim\,$1
per square degree requires an angular resolution of $\sim\,$1$'$, and
the capability to separate brightness fluctuations of size
$\sim\,$10$\,{\rm mJy}$ (for 150 GHz) from the primary CMB
fluctuations. Several current experiments are capable of meeting such
requirements, for example the OVRO and BIMA interferometers, and they
are already being used to detect and study SZ clusters. Improvements
to these experiments should soon reduce the upper bounds shown in
Fig.~\ref{fig:lcl} and make the thermal SZ signal we predict
detectable in the near future. The planned satellite mission {\em
Planck Surveyor} with its high sensitivity and multi-frequency
capability should allow a clean separation of the thermal SZ signal
from the primary anisotropies, even though its angular resolution will
be limited to less than $\ell$$\,\sim\,$1000.  However, future
interferometric missions at mm-wavelengths, like ALMA, will be able to
probe much smaller angular scales, up to $\ell$$\,\sim\,$$10^6$, and
should have enough sensitivity to detect both the thermal and kinetic
effects. More detailed discussions of the prospects for extracting the
SZ signal from upcoming CMB experiments are given elsewhere (Cooray,
Hu \& Tegmark \cite{Cooray2000}, Holder \& Carlstrom~\cite{Hold1999},
and Bartlett~\cite{Bart2000}, among others).

We have examined the dependence of the SZ signal on early energy
injection using an ``extreme'' model where $0.1\,$keV to $1\,$keV of
energy was added into all of the gas at $z=7$.  The energy injection
heats the gas, but also drives it out of small halos, erasing
small-scale structure.  The loss of small scale structures causes a
reduction of the angular power spectrum at high-$\ell$ while the
heating of the gas enhances the large angle signal (see
Fig.~\ref{fig:powerinj}).

In the runs without energy injection, most of the mean Comptonization
is generated at redshifts $z \le 3$, with a small tail extending to a
redshift of about 5 (da Silva~et al.~\cite{SBLT}).  However, in the
runs with energy injection, additional Comptonization is provided by
hot ionized gas in the IGM at still higher redshift.  Current FIRAS
limits on the mean Comptonization already exclude our most extreme
model that injected $1\,$keV at $z=7$.  In future work, we will
investigate the dependence of the SZ effects on additional input
physics such as radiative cooling, star formation and its associated
energy input by supernova feedback.

\bigskip
\acknowledgments We would like to acknowledge useful conversations
with Rupert Croft, Chris Metzler, and Bill Holzapfel.  This work was
supported by NASA Astrophysical Theory Grant NAG5-3820 and by the NSF
under grants ASC93-18185, ACI96-19019, and AST-9802568.  M.W. was
supported by NSF-9802362 and a Sloan Fellowship.  The simulations were
performed at the San Diego Supercomputer Center.

\end{document}